\documentclass[twocolumn,floatfix]{revtex4}
\usepackage{epsf,graphicx}
\usepackage{amsmath,amssymb,amsfonts}
\newcommand{\lora} {\boldmath$\longrightarrow$}
\newcommand{\vecbm}[1]{\mbox{\boldmath#1}}

\newcommand{\lola} {\boldmath$\longleftarrow$}

\newcommand{\nvec}[1]{\stackrel{\rightarrow}{#1}}

\begin{document}

\title{Challenges about entropy.}
\author{D.H.E. Gross}
\affiliation{Hahn-Meitner Institut and Freie Universit{\"a}t Berlin,
Fachbereich Physik, Glienickerstr. 100, 14109 Berlin, Germany,
\\gross@hmi.de; WEB-page: http://www.hmi.de/people/gross/}

\begin{abstract}
The physical meaning of entropy is analyzed in the context of
statistical, nuclear, atomic physics and cosmology. Only the
microcanonical Boltzmann entropy leads to no contradictions in
several simple, elementary and for thermodynamics important
situations. The conventional canonical statistics implies several
serious errors and misinterpretations. This has far reaching
consequences for phase-separations as well for the usual
formulations of the second law. Applications in cosmology suffer
under the ubiquitous use of canonical statistics. New
reformulations in terms of microcanonical statistics are highly
demanded but certainly difficult.

\end{abstract}
\maketitle

\section{Introduction}
Entropy, S, is the characteristic entity of thermodynamics and
thermostatistics \cite{clausius1865}. It distinguishes thermodynamics from
all other physics; therefore, its proper understanding is essential.
However, this is sometimes obscured by frequent use of the Boltzmann-Gibbs
canonical ensemble, and the thermodynamic limit. Also its relationship to
the second law is often beset with confusion between external transfers of
entropy $d_{ext}S$ and its internal production $d_{int}S$. The main source
of the confusion is of course the lack of a clear {\em microscopic and
mechanical} understanding of the fundamental quantities of thermodynamics
like heat, external vs. internal work, temperature, and last not least
entropy, at the times of Clausius and possibly even today. I got some
encouragement from reading Redlich's critique \cite{redlich68}.

The second law of the necessary increase (or constancy) of the
entropy has far reaching significance for all physics and
especially for astro-physics and cosmology. This demands a new
critical discussion of the basics of statistical mechanics and the
use of the various ensembles.

This contribution wants to give a first glance of these
challenges. Of course it is not able to go into the details. This
would demand a whole course in statistical physics. The author's
work on statistical mechanics is given by \cite{gross174} and some
recent ones \cite{gross216,gross220}. His interpretation of the
foundation of entropy and the second law  is given in
refs.\cite{gross207,gross212,gross184,gross183}

The boxes show the transparencies of  my talk presented at the
XLIV  international winter meeting on nuclear physics in Bormio
(Italy) - 2006, January 29- February 5. I add a few more
explanations.

\begin{center}
\fbox{\begin{minipage}{8.5cm}
\begin{center}
{\bf Plan of the talk}\\~\\ Entropy by Clausius
\\~\\Entropy by Boltzmann\\~\\ Caratheodory\\~\\Entropy by Prigogine\\~\\Entropy by
Tsallis\\~\\ Phase transitons\\~\\Convexities of S(E)\\~\\
Long-range interactions, gravity\\~\\Arrows of time\\~\\
Cosmological perspectives and speculations
\end{center}
\end{minipage}}
\fbox{\begin{minipage}{8.5cm}
\begin{center}
{\bf Clausius}\cite{clausius1865}:
\\~\\ Temperature = unambiguous control parameter of
thermodynamics
\begin{eqnarray*}
dS&=&\frac{dQ}{T}\\ \lefteqn{\hspace{-2cm}
 \mbox{``value of metamorphosis''}}\\
 \lefteqn{\hspace{-1cm}\mbox{ or ENTROPY}}\\~\\
\lefteqn{\hspace{-1cm}\mbox{SECOND LAW
:}}\\\lefteqn{\hspace{-2.5cm}\mbox{``Heat only flows from hot to
cold''}}
\\~\\\Downarrow\\~\\ \oint
{\frac{dQ}{T}}&\ge&0\\
 \lefteqn{\hspace{-3cm}\mbox{= uncompensated metamorphosis}}
\end{eqnarray*}
\end{center}
\end{minipage}}

\fbox{\begin{minipage}{8.5cm}
\begin{center}
{\bf Boltzmann-Planck :}\\
\begin{equation} \fbox{\fbox{\vecbm{S=k*lnW}}}\label{boltzmann0}
\end{equation}as written on Boltzmann's tomb-stone, with
\begin{center}
\begin{eqnarray}
W(E,N,V)&=&\nonumber\\ \lefteqn{\hspace{-3cm}
\int{\frac{d^{3N}\nvec{p}\;d^{3N}\nvec{q}}{N!(2\pi\hbar)^{3N}}
\epsilon_0\;\delta(E-H\{\nvec{q},\nvec{p}\})}}\label{boltzmann}
\end{eqnarray}
\end{center}
 =number of initial states with\\ the same macroscopic
constraints
\\ $\equiv$ microscopic ignorance\\~\\ $\Rightarrow$
\underline{no thermodynamic limit}\\no extensivity\\no concavity\\no
homogeneity.\\\underline{no Boltzmann-Gibbs
(canonicality)\cite{gross216}}\\no Caratheodory\\no Tsallis \cite{gross203}

\end{center}
\end{minipage}}
\end{center}

The Boltzmann-Planck formula has a simple but deep physical interpretation:
$W$ or $S$  measure our ignorance about the complete set of initial values
for all $6N$ microscopic degrees of freedom which are needed to specify the
$N$-body system unambiguously, ref.\cite{kilpatrick67}. To have complete
knowledge of the system we would need to know [within its semiclassical
approximation (\ref{boltzmann})] the initial positions and velocities of
all $N$ particles in the system, which means we would need to know a total
of $6N$ values. Then $W$ would be equal to one and the entropy, $S$, would
be zero. However, we usually only know the values of a few macroscopic
parameters that are conserved or change slowly with time, such as the
energy, number of particles, volume and so on. We generally know very
little about the positions and velocities of all the particles. The
manifold of all these points in the $6N$-dim. phase space, consistent with
the given conserved macroscopic constraints of $E,N,V,\cdots$, is the
microcanonical ensemble, which has a well-defined geometrical size $W$ and,
by equation (\ref{boltzmann0}), a non-vanishing entropy, $S(E,N,V,\cdots)$.
The dependence of $S(E,N,V,\cdots)$ on its arguments determines completely
thermostatics and equilibrium thermodynamics.

Clearly, Hamiltonian (Liouvillean) dynamics of the system cannot create the
missing information about the initial values - i.e. the entropy
$S(E,N,V,\cdots)$ cannot decrease. As has been further worked out in
ref.\cite{gross183} and more recently in ref.\cite{gross207} the inherent
finite resolution of the macroscopic description implies an {\em increase}
of $W$ or $S$ with time when an external constraint is relaxed. Such is a
statement of the second law of thermodynamics, ref.\cite{prigogine71},
which requires that the {\em internal} production of entropy be positive or
zero for every spontaneous process. Analysis of the consequences of the
second law by the microcanonical ensemble is appropriate because, in an
isolated system (which is the one relevant for the microcanonical
ensemble), the changes in total entropy must represent the {\em internal}
production of entropy, see below, and there are no additional uncontrolled
fluctuating energy exchanges with the environment.

\section{Second Law}

\begin{center}
\fbox{\begin{minipage}{8.5cm}
\begin{center}
{\bf Second Law}\\~\\
 P. Glansdorf and I. Prigogine \cite{prigogine71}\\~\\
\begin{eqnarray} dS &=&d_{ext}S + d_{int}S \nonumber\\ d_{ext}S&
=&\mbox{positive, negative, or zero} \nonumber\\ d_{int}S &\ge& 0
\mbox{=Second Law}\label{prig}
\end{eqnarray}~\\
Thus the Second Law\\ is explicitly and most clearly defined by\\
MICROCANONICAL STATISTICS !
\end{center}
\end{minipage}}
\end{center}
From our experience (see section(\ref{phtr})) we know that the fundamental
microcanonical ensemble is {\em not} equivalent to the conventional
canonical ensemble. This is the case just in the thermodynamically most
interesting situations: at phase separation see section(\ref{phtr}). In
order to have a correct formulation of the second law we must follow
Boltzmann's or Prigogine's \cite{prigogine71} formulation of microcanonical
entropy.

Already Clausius himself gave an illuminating example in
\cite{clausius1854} that clearly shows how the conventional
(canonical) formula:
\begin{equation}
dS=\frac{dE}{T} \label{entropy}
\end{equation}
fails and is {\em not} automatically the correct expression of the
entropy change:

When an ideal gas suddenly streams under insulating conditions from a small
vessel with volume $V_1$ into a larger one ($V_2>V_1$), neither its
internal energy $U$ or $E$, nor its temperature changes, nor external work
done, but its internal (Boltzmann-)entropy $S_{int}$ e.q.(\ref{boltzmann0})
with (\ref{boltzmann}) rises, by $\Delta S=N\ln{(V_2/V_1)}$. This is also a
clear example for a microcanonical situation where the entropy change by an
irreversible metamorphosis of the system is absolutely internal. It has
{\em nothing} to do with heat exchange with the surrounding as expressed by
eq.\ref{entropy}. In this context see also my controversy with Jarzynski
\cite{gross218}.

Then the second law becomes very simple and claims the increasing loss of
information (rise of entropy) due the impossibility to distinguish the
highly ``fractal'' distribution in phase-space after some mixing dynamics
from its surrounding. I. e. the internal entropy {\'a} la Prigogine
(\cite{prigogine71}) of the necessarily redundant macroscopic description
of the system becomes larger (information-loss)\cite{gross183}.

This example demonstrates the problematic of the canonical description of
thermo-statistics. This is even more evident when the original object of
thermodynamics, the description of boiling water, is addressed. This is
discussed in the next section.
\begin{center}
\begin{widetext}\rule{-1mm}{-2cm}
\section{Phase transitions \label{phtr}}
 \fbox{\begin{minipage}{18cm}
\begin{center}
{\bf Phase-transitions in conventional (grand-canonical)
statistics in the thermodynamic limit:\\ Yang-Lee singularities.
Why?}\\~\\
 $\Rightarrow$ Only homogeneous configuration$\Leftarrow$\\ $\Rightarrow$ No boiling water!$\Leftarrow$
\\ Are phase transitions really so trivial ???
\\~\\ Physical reason:\\~\\
\lora Strong inhomogeneities\lola,\\ \lora Internal surfaces
\lola\\~\\ What is the difference between\\~\\
{\Large\bf\boldmath\begin{tabular}{c||c||c}
Solid&Liquid&Gas~~~?\\\hline\hline \multicolumn{2}{c||}{have
surfaces}&no surface\\ \multicolumn{2}{c||}{condensate volume less
than external volume }&fills any volume\\ surface has
edges&surface has no edges\\hard fixed&flexible, adjusts to
container&
\end{tabular}}\\~\\$\Rightarrow$ Moral: Avoid thermodynamic limit
!$\Leftarrow$
\end{center}
\end{minipage}}
\end{widetext}

\fbox{\begin{minipage}{8.5cm}
\begin{center}

{\LARGE  BUT}\\
    Without thermodynamic limit several gospels of conventional
    thermodynamics become obsolete:
\begin{itemize}
\item Equivalence of canonical and microcanonical ensembles
\item Legendre transform structure
\item Phase transitions exist only in the thermodynamic limit
\item Specific heat is $\sim <\!\!(\delta E)^2\!\!>\;\; >0$
\item Extensivity of $S$, scaling with $N$
\item Concavity of $S(E,N)$
\item Rise of entropy is connected to trend towards uniformization
\item Second Law only in infinite systems
\end{itemize}
\end{center}\rule{0mm}{0.5cm}
\end{minipage}}
\begin{widetext}
\fbox{\begin{minipage}{18cm}
\begin{center}
{\bf Convexity of S(E)}\\~\\ Example:\\ Atomic cluster fragmentation gives
\\ detailed insight into region\\ of phase-separation,
\underline{here no Coulomb}
\\ $\mapsto$ most interesting physics
\includegraphics*[bb = 11 1 491 541,
angle=-90, width=17cm, clip=tru]{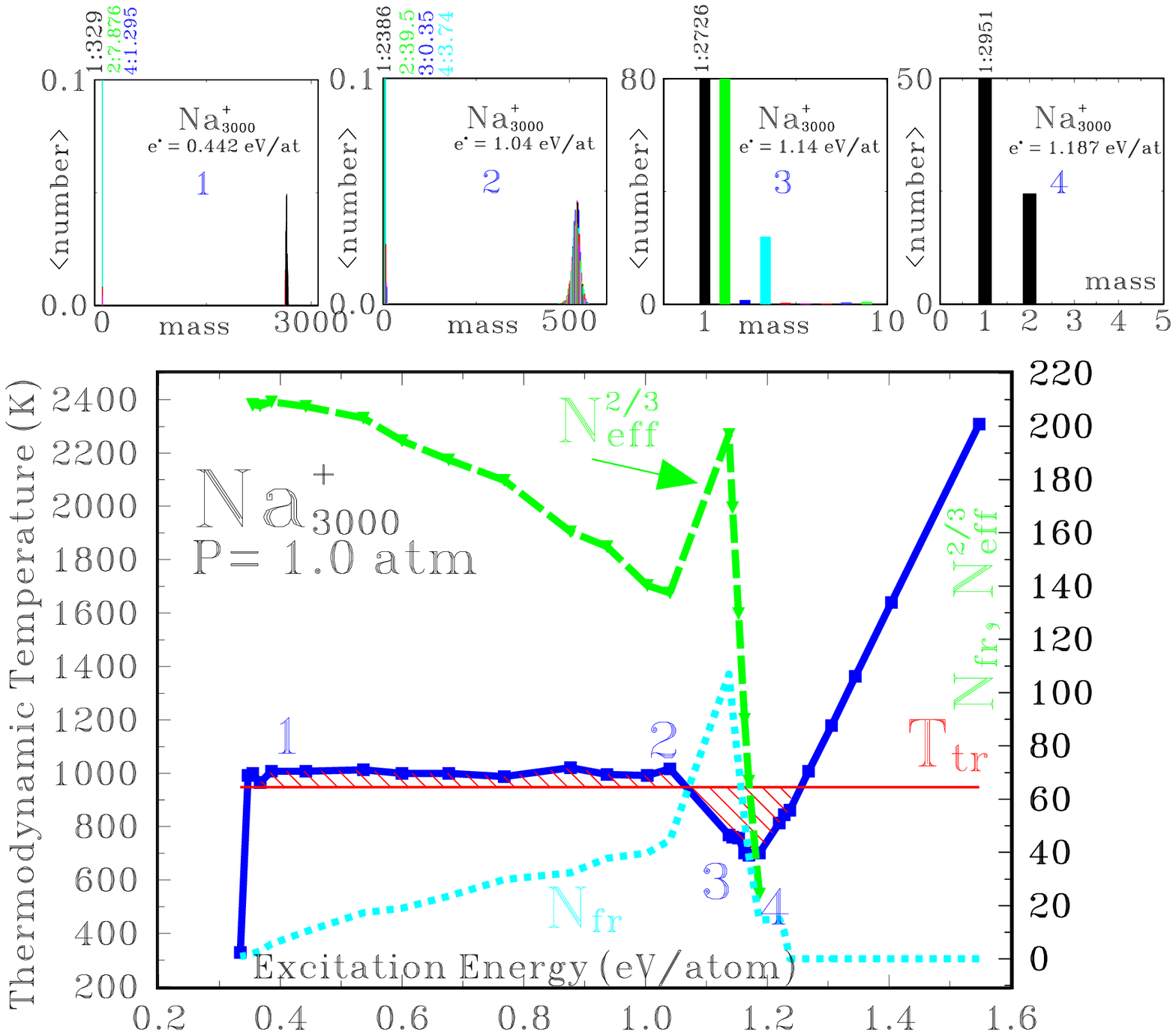}\\ {\bf``Compound nucleus
ever''\cite{bowman87}?}
\end{center}
Caloric curve of $3000$ Na atoms at a pressure of $1$ atm with
realistic interaction (c.f. \cite{gross174}). The temperature $T$
in Kelvin, the energy in eV/atom. $N_{fr}$ the number of
fragments, $N^{2/3}_{eff}=\sum{N_i^{2/3}}$ the number of surface
atoms of dimers and heavier ones. The four inserts give the
fragment distribution at 4 characteristic energies over the
intruder. E.g.: at $0.442$ eV 1:329 means 329 monomers, and
2:7.876 means 7.876 dimers,and 4:1.295 means 1.295 quadrimers on
average. Below $e\sim 0.33$ eV $2999$ atoms are condensed in a
single liquid droplet and above $e\sim 1.3$ eV nearly all atoms
are free ($\sim$ ideal gas). Color online

\end{minipage}}

\end{widetext}
\end{center}
The above example of the liquid-gas transition of first order in a
mesoscopic droplet clarifies the errors made by the recent works
by \cite{bowman87}: Whereas the transition extends clearly over
the latent heat from $\propto 0.3$ to $\propto 1.3$ electron volt,
the transition temperature is {\em below} the evaporation line
from point $1$ to $2$. This is the region of evaporation
configuration as is clearly seen by the inserts. The temperature
of this piece remains above the true transition temperature
$T_{tr}$. This is of course because phase separation is a
microcanonical phenomenon characterized by a {\em back-bending}
caloric curve, which does not exist in the canonical ensemble
\cite{gross220}. The final backbending is clearly seen beyond
point 2. One might argue: the system is unstable in the
backbending region. This is wrong as under a control of the energy
fluctuations (thermos-flask, large systems like the ocean). Then
the intermediate region is very well accessible.

As seen in the mass distribution of the fragments (inserts) the
quality of the distribution changes over the latent heat  from
evaporation-like (insert 1 to insert 2, called by Bowman et al.
\cite{bowman87} in a somewhat premature way ``compound nucleus for
ever'') finally to multifragmentation at 3 and 4. As function of
the {\em dynamically conserved} energy the weight $e^{S(E)-E/T}$
of the configurations with energy E in the definition of the
canonical partition sum
\begin{equation*}
Z(T)=\int_0^\infty{e^{S(E)-E/T}\frac{dE}{\epsilon_0}}\label{canonicweight}
\end{equation*} becomes here {\em bimodal over the range of the latent
heat.}

Thus, in contrast to what was claimed at this meeting, this
bimodality over the latent heat is a {\em necessary signal of any
phase separation}. In {\em nuclear} fragmentation the caloric
curve $T(E)$ is often dramatically modified by additional fission
decays \cite{gross174}.

\section{Cosmological challenges}
In very recent times possible links between general relativity and
statistical mechanics became a topic of hot speculations.

First the philosophical question whether the arrow of time as expressed by
the second law of thermodynamics has anything to do with the development of
the universe (cosmological arrow of time) was a topic already of Clausius
\cite{clausius1865}. Recently Allahverdian and Gurzadyan discuss this in
\cite{allahverdian02}. Uffink's \cite{uffink01} answer to this question is:
``The arrow of time has nothing to do with the second law!'' This shows the
wide span of controversial  opinions about this question.

Allahverdian and Gurzadyan \cite{allahverdian02} assume that the system is
coupled to a memory-loosing bath. I.e. the time evolution operator of the
bath
\begin{enumerate}
\item $\mathcal{T}(t,0)=\mathcal{T}(t)$ after a sufficiently long time $t$,
which must still be short compared with the Poincar{\'e}
recurrence time.
\item and the system relaxes to the stationary distribution
$\mathcal{D}_S(t)\rightarrow \mathcal{D}_S^{(st)}$.
\end{enumerate}
Both are of of course strong assumptions which imply the second
law.

The ``ensemble probabilistic'' interpretation of statistical
mechanics \cite{gross183} takes the fact seriously that knowing
only a few macroscopic parameters of the system leaves the
majority of degrees of freedom uncontrolled. Therefore this fixes
only the ensemble of many microscopic realizations of the system.
Thermodynamics as also thermo-statistics can only describe the
average (over the ensemble) behavior. Certainly there is nothing
like a single ``macroscopic state'' with a discrete Poincar{\'e}
recurrence time. In general the whole ensemble will never recur.

This, however, seems to have nothing to do with a cosmological arrow of
time and in contrast to Allahverdian and Gurzadyan \cite{allahverdian02} I
tend to agree with Uffink \cite{uffink01} that it might be an advantage to
abandon the idea of a relation of the second law  with the arrow of time.
\begin{center}
\fbox{\begin{minipage}{8.5cm}
\begin{center}
{\bf Different arrows of time:}\\~\\ Allahverdyan and Gurzadyan
\cite{allahverdian02}:\\
\begin{enumerate}
\item Thermodynamic arrow of time:\\
second law
\item Cosmological arrow of time:\\
universe expands
\item Psychological arrow of time:\\
we only know about the past, not future\\ we become older and more
stupid
\item Electromagnetic:\\retarded interaction
\item Quantum-mechanical:\\
wave-function during measurement collapses
\end{enumerate}
~\\~\\$1 \rightleftarrows 2$ ? \\~\\

$\Rightarrow$ Deeper understanding of entropy !$ \Leftarrow$
\end{center}
\end{minipage}}
\fbox{\begin{minipage}{8.5cm}
\begin{center}
{\bf Cosmological perspectives and speculations}\\~\\
 Common wisdom:\\ Specific heat of self-gravitating system often
 negative.\\~\\ However, canonically:
\begin{equation*}
C(T) =\frac{<(\Delta E)^2>}{T^2}\ge 0
\end{equation*}
$\Rightarrow$ gravitating systems are not canonical $\Leftarrow$
\end{center}
\end{minipage}}
\fbox{\begin{minipage}{8.5cm}
\begin{center}
{\bf Black Hole :}
\begin{eqnarray}
\lefteqn{\mbox{e.g.: Schwarzschild horizon:}}\nonumber\\
  R&=&2M =2E\nonumber\\
A&=&16\pi M^2\nonumber\\ dA&\ge 0\nonumber\\\sim
dS&\ge&0\nonumber\\~\nonumber\\
\lefteqn{\mbox{\hspace{-1cm}Hawking radiation, thermal
spectrum:}}\nonumber\\T=\frac{1}{8\pi M}&=&\frac{1}{8\pi
E}\nonumber\\~\nonumber\\
 \lefteqn{\mbox{negative specific heat !}}\nonumber\\
 &\Downarrow&\nonumber\\ S_{BH}&=&A/4=\pi R^2=4\pi M^2\nonumber\\~\nonumber\\
\lefteqn{\hspace{-1cm}\Rightarrow\mbox{e.g.: Generalized Second
Law:}}\nonumber\\ S_{initial}&=&S_{BH}(M)+S_{matter}\nonumber\\
&\le& S_{BH}(M+M_{matter})\label{genseclaw}
\end{eqnarray}
 \\$\Rightarrow$ Limits the number in $\frac{4\pi}{3}R^3$ of\\
 different elementary fields with $\lambda \le R$\\When too many they disappear in
 black hole!
\end{center}
\end{minipage}}
\end{center}

This shows the challenge of a possible connection between entropy and the
horizon of black holes \cite{bekenstein73,bousso02}.

Explicitly the generalized second law eq.(\ref{genseclaw}) sets
automatically an upper end to the spectrum of any system. $S(E)$ is
necessarily overall concave. If the energy of a system becomes too large it
becomes so heavy that it implodes into a black hole. This also illuminates
the absurdity of the Hagedorn constant temperature spectrum as recently
discussed and beautifully demonstrated by Moretto
\cite{moretto05,moretto06}.
\section{Conclusion}
The lesson I like to transfer here is that the canonical statistics is
sometimes dangerously erroneous and must then be replaced by the
microcanonical statistics. However, nearly all works on cosmology ignore
this fact. Certainly, in general relativity a lot more work must be done
along this challenge \cite{gross220}. Examples showing some of the
perspectives are \cite{bousso02,padmanabhan06,eling2006}.

In \cite{bousso02} is the remarkable sentence: ``Entropy costs energy and
energy focuses light into a caustics of space-time. The caustics prevent
light-sheets going on for ever.'' The finite size of light-sheets and the
surface-area of the corresponding system determines entropy. This sentence
characterizes the deep link between entropy and space-time and the specific
significance of the microcanonical statistics in contrast to any intensive
(canonical) description. There is a further important statement in
\cite{bousso02}: ``Entropy is a non-local phenomenon. Only in some
approximation entropy can be described by a local flow of entropy
density'', i.e. by an intensive (canonical) description. It is fascinating
how the experience from ordinary condensed matter and especially
phase-transitions points to the same conclusion as cosmology: Entropy is
non-local and depends primarily on the {\em ``extensive''} control
parameters as energy. One may argue in view of the large scales of the
universe, a local {\em canonical} approximation might be acceptable.
However, gravity has also a long range. The ubiquitous negative heat
capacity of self-gravitating system should be a warning.




\end{document}